  \documentclass[twocolumn]{igs}
  \usepackage{igsnatbib}
  \usepackage{stfloats}
  \usepackage{amsmath}
  \usepackage{lineno}
  \ifx\pdftexversion\undefined
    \usepackage[dvips]{graphicx}
  \else
    \usepackage[pdftex]{graphicx}
  \fi
  \usepackage{caption}
  \usepackage{subcaption}
  
\begin{document}

\title[Ross Ice Shelf Radio Properties]{Radar Absorption, Basal Reflection, Thickness, and Polarization Measurements from the Ross Ice Shelf}

\author[J.C. Hanson and others]{Jordan C. HANSON,$^{1,2}$\protect\thanks{Present address:
  4073 Malott Hall, Dept. of Physics and Astronomy, University of Kansas, Lawrence, KS 66045, USA. j529h838@ku.edu} ~Steven W. BARWICK,$^1$ Eric C. BERG,$^1$ Dave Z. BESSON,$^{2,5}$ Thorin J. DUFFIN,$^{1}$ Spencer R. KLEIN,$^4$ 
  Stuart A. KLEINFELDER,$^3$ Corey REED,$^1$ Mahshid ROUMI,$^1$ Thorsten STEZELBERGER,$^4$ Joulien TATAR,$^1$ 
  James A. WALKER,$^1$ Liang ZOU$^1$}

\affiliation{
$^1$Dept. of Physics and Astronomy, University of California, Irvine, Irvine,
  CA 92697, USA. \\
$^2$Dept. of Physics and Astronomy, University of Kansas, Lawrence, KS 66045,
	USA. \\
$^3$Dept. of Electrical Engineering and Computer Science, University of of California,
	Irvine, CA 92697, USA.\\
$^4$ Lawrence Berkeley National Laboratory, Berkeley, CA 94720, USA.\\
$^5$ Moscow Physics and Engineering Institute, Moscow, 115409, Russian Federation, Russia.}

\abstract{Radio-glaciological parameters from Moore's Bay, in the Ross Ice Shelf, have been measured.  The thickness of the ice shelf in Moore's Bay was measured from reflection times of radio-frequency pulses propagating vertically through the shelf and reflecting from the ocean, and is found to be $576\pm8$ m.  Introducing a baseline of 543$\pm$7 m between radio transmitter and receiver allowed the computation of the basal reflection coefficient, $R$, separately from englacial loss.  The depth-averaged attenuation length of the ice column, $\langle L \rangle$ is shown to depend linearly on frequency.  The best fit (95\% confidence level) is $\langle L(\nu) \rangle = (460\pm20)-(180\pm40)\nu$ m (20 dB/km), for the frequencies $\nu=$[0.100-0.850] GHz, assuming no reflection loss.  The mean electric-field reflection coefficient is $\sqrt{R}=0.82\pm0.07$ (-1.7 dB reflection loss) across [0.100-0.850] GHz, and is used to correct the attenuation length.  Finally, the reflected power rotated into the orthogonal antenna polarization is less than 5\% below 0.400 GHz, compatible with air propagation.  The results imply that Moore's Bay serves as an appropriate medium for the ARIANNA high energy neutrino detector.}

\maketitle

\section{Introduction}

The vast ice sheets of Antarctica have become important to high-energy neutrino physics in recent years \citep{RICE,IceCube,ANITA,Barwick,review}, motivated by the convenient properties of glacial ice, including optical and radio-frequency dielectric properties.  High-energy cascades induced by neutrinos emit Cherenkov photons; photons with 350-500 nm wavelengths can propagate 10-100 m in Antarctic ice before being detected by photomultiplier tubes \citep{IceDust}.  Similarly, at energies $\gtrsim 0.1$ EeV, neutrinos begin to produce measurable $\emph{Askaryan}$ pulses \citep{Askaryan}, a form of coherent Cherenkov radiation in the radio-frequency (RF) regime.  Moore's Bay, part of the Ross Ice Shelf, presents an attractive target volume for studying these particle interactions, because the radiation experiences minimal attenuation in the cold ice, and can be reflected back towards the surface by the ocean.  Knowledge of the dielectric properties of the ice shelf is required to build such an experiment.

By transmitting a $\sim 1$ ns pulse through a transmitting antenna downwards through the ice shelf, and recording the reflections from the oceanic interface, bulk ice attenuation and the reflection coefficient of the interface can be inferred.  Using reflected radio pulses to study ice sheets and shelves is known as radio-echo sounding, and it has been used to study glaciers and ice shelves in various locations on the Ross Ice Shelf (RIS) and the high plateau \citep{Neal79,Neal82,Besson,Cresis,ARA}.

Basal reflection in Moore's Bay has been studied previously.  C.S. Neal \citep{Neal79,Neal82} reported on the RIS, using a 60 ns wide, 60 MHz pulse, recording the returned power versus location.  Flights 1 km above the RIS were performed, including several points over Moore's Bay.  Basal reflection coefficients were derived in 10 dB increments, assuming no losses from dust or other impurities, for contours across the shelf.  Moore's Bay produces reflection coefficients near the Fresnel limit (-0.82 dB, or $\approx 0.91$ for the electric field), and two explanations were offered.  Moore's Bay is far from brine percolation zones that are traced from grounding line to the shelf front, which are correlated with ice velocity.  Second, the melt rate near the grounding line for basal ice would prevent the formation of an abrupt basal layer of saline ice, and instead replace glacial ice with saline ice over time.  The freeze-on of saline ice at the shelf bottom does occur, however these regions are far from the location of ARIANNA, and the average accumulation rate of bottom saline ice is only $0.3\pm0.1$ m/yr in the East RIS \citep{Rignot}.

C.S. Neal has shown that two parameters besides peak power can be extracted from the data \citep{Neal82}.  The width of the peak power distribution for a specific location pertains to vertical roughness at the oceanic interface.  Second, the spatial correlation of power measurements reveal horizontal correlation lengths for roughness.  These measurements must be compared to theoretical distributions of the same parameters, from the theory of rough-surface scattering \citep{Kam}.  The most general statistical surface, with the fewest number of parameters, was chosen: a Gaussian surface roughness described by normal fluctuations about a mean depth, and a specified horizontal correlation length.  C.S. Neal reports vertical rms of 3 cm at the ocean/ice boundary, spread over correlation lengths of 27.5 m, thirty kilometers east of Ross Island.  The results make no use of absolute power measurements, and thus are independent of assumptions of RF absorption.

\begin{figure}
\begin{center}
\includegraphics[trim=4mm 4mm 4mm 0mm,clip=true,width=80mm]{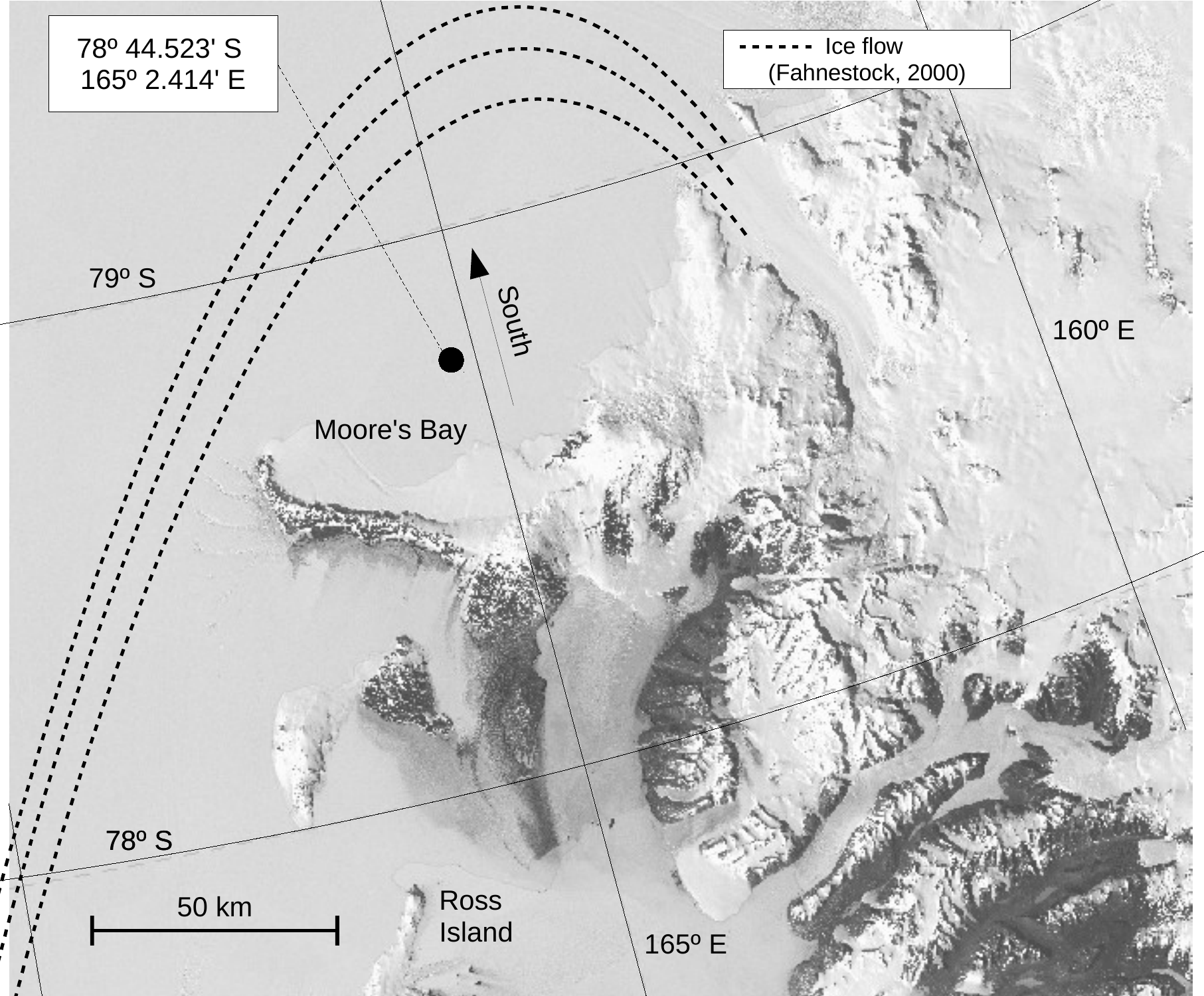}
\end{center}
\caption{\label{map} The site studied in this work is marked with the black circle.  Moore's Bay is the area south of Ross Island, enclosed by Minna Bluff.  The satellite data is made available by the US Geological Survey \citep{LIMA}.  The main ice flow lines are illustrated with dashed lines.  For further analysis and discussion, see \citep{Fahnestock}.}
\end{figure}

To measure the reflection coefficient separately from englacial loss, the transmitting and receiving antennas may be separated by a baseline comparable to the shelf-depth.  The signal path is longer with the baseline, providing different absorption, but the same reflection loss \citep{Hanson2011,Jordo}.  By comparing total loss in the set-up with and without a baseline, absorption and reflection loss can be measured separately.  A map of the site studied in this work is shown in Fig. \ref{map}.

\subsection{RF Attenuation Length}\label{sec:Lambda}

The amplitude of an electric field decreases by $1/e$ after propagating one attenuation length.  For an electromagnetic plane wave travelling through a dielectric medium with a complex index of refraction $n = n'-in''$, the electric field is:

\begin{equation}
\textbf{E} = \textbf{E}_0 \exp(-i(nkx-\omega t)) = \textbf{E}_0' \exp \lbrace -n''kx \rbrace
\end{equation}

The electric field attenuation length is then $(n''k)^{-1} = L$ ($\omega$ is the angular frequency).  When measured over a volume of material with varying dielectric absorption, the attenuation length is averaged over the effect of depth on the dielectric constant $\epsilon = \epsilon'-i\epsilon''$, and in turn the $\emph{loss tangent}$, $\tan \delta = \epsilon''/\epsilon'$.  If $\tan \delta \ll 1$, it can be shown that

\begin{align}
\langle L \rangle^{-1} &= (\pi \nu / c) \sqrt{\epsilon'} \tan \delta ~~ (m^{-1})\\
N_L (dB/km) &= 8686.0 \langle L[m] \rangle^{-1} \label{stupidAssReviewWhoDoesntHaveACalculator}
\end{align}

Equation \ref{stupidAssReviewWhoDoesntHaveACalculator} is the conversion from attenuation length to absorption loss, $N_L$, in dB/km.  The $\emph{Debye}$ model shows that $\nu \tan \delta$ is approximately constant, provided the frequency is far from any molecular resonances (this is true for 0.1-1 GHz).  Additionally, $\epsilon'$ (in ice) is constant for the bandwidth 0.1-2 GHz.  Thus, frequency-dependence in $\langle L \rangle$ is attributed to other effects, such as acids and sea-salt impurities \citep{Bogorodsky,Mat2}.

\subsection{Reflection coefficient}\label{sec:Refl}

Under the Debye model, with a single relaxation time, the ice conductivity is $\sigma = 2 \pi \epsilon_0 \epsilon' \nu \tan \delta \approx 10~\mu$S/m at 100 MHz \citep{DandE2004}.  By comparison, sea water has a conductivity of a few S/m, with a skin depth of 30 mm, at 60 MHz \citep{DandE2004,Somaraju}.  The reflection coefficient for the electric fields ($\sqrt{R}$, where $R$ refers to power) is given by $(n_1-n_2)/(n_1+n_2)$, given the complex $n_1 = \sqrt{\epsilon_1}$ and $n_2 = \sqrt{\epsilon_2}$ for the dielectric and conductive media, respectively.

\begin{align}
\begin{split}
&\lim_{\tan{\delta_2} \gg 1, \tan{\delta_1} \rightarrow 0}  | \sqrt{R} | = \left \lvert \frac{1-n_2/n_1}{1+n_2/n_1} \right \rvert \\
&= \left \lvert \frac{1-\sqrt{\alpha}e^{-i\delta_2/2}}{1+\sqrt{\alpha}e^{-i\delta_2/2}} \right \rvert =\left ( \frac{1+\alpha-\sqrt{2\alpha}}{1+\alpha + \sqrt{2\alpha}} \right )^{1/2}
\label{R3}
\end{split}
\end{align}

Equation \ref{R3} demonstrates that $|\sqrt{R}| \to 1$, where $n_1$ refers to the ice and $n_2$ refers to the ocean, given the limits $\tan\delta_2 \gg 1$, and $\tan\delta_1 \approx 0$, and $\alpha = \epsilon_2''/\epsilon_1'$.  In Eq. \ref{R3}, the fact that $\delta_2 \rightarrow \pi/2$ has been used.  Equation \ref{R3} is completely general as long as the limit is satisfied.  The right-hand side has a global minimum at $\alpha=1$, or $\epsilon_2'' = \epsilon_1'$, corresponding to a minimum electric field reflection coefficient of $\sqrt{R}_{min}\approx$0.41.  Realistic values for both ice and sea water indicate $\alpha$ ranges from 20 to 30, depending on the salinity and temperature of the sea water \citep{Somaraju,DandE2004}.  C.S. Neal \citep{Neal79} suggested that the reflection coefficient in Moore's Bay is $\approx -0.82$ dB, or $\sqrt{R}=0.91$, based on the properties of the sea-water beneath the RIS.  These upper and lower bounds form an allowed range of $\sqrt{R} = 0.41-0.91$.

In addition to vertical radio echoes, measurements were taken with a baseline distance between transmitter and receiver, introducing a new overall path length.  In this work, these measurements are named $\emph{angled}$ $\emph{bounce}$ studies.  For the angled bounce studies reported here, Eqn. \ref{secOrder} shows that the reflected power limits to the expression for normal incidence (for s-polarized waves).  Also, the initial transmission angle from normal is reduced, because the upper firn layer bends the transmitted pulse downward (to $\theta\lesssim 30^{\circ}$), given the initial antenna orientation of $45^{\circ}$.  Ignoring the cosine dependence in Eq. \ref{secOrder} amounts to a 1-5\% correction, depending on $n_2$.

\begin{equation}
\sqrt{R} \approx \frac{n_1(1-\theta^2/2)-n_2(1-\frac{1}{2}(\epsilon \theta)^2)}{n_1(1-\theta^2/2)+n_2(1-\frac{1}{2}(\epsilon \theta)^2)} \approx \frac{n_1-n_2}{n_1+n_2} \label{secOrder}
\end{equation}

\subsection{Ice Thickness Calculation}\label{sec:Depth}

The upper 60-70 m of the ice shelf is a mixture of snow and ice known as firn, which has a density of $\approx 0.4$ g/cc near the surface \citep{ARIANNANIM}.  This result is in agreement with the value 0.36 g/cc from \citep{DandE2004}.  Looyenga's equation, $n_{ice}$, and the firn surface density predict the firn index to be $n_{firn}\approx1.3$.  This number was confirmed with pulse propagation timing at the surface, over a distance of $543\pm7$ m (see below for detail).  From the pulse arrival time, the implied wave speed indicated an index of $n_{surf}=1.29\pm0.02$ \citep{Jordo}.

The density and thus the index of refraction has an exponential depth dependence, according to the Schytt model:

\begin{align}
n(z) &= 1.78 = n ~~ (z\geq 67 m)  \label{nz1}  \\
n(z) &= n_0 + p \exp(-z/q) ~~(z<67 m)  \label{nz2} 
\end{align}

In Eq. \ref{nz2}, $n_0=1.86$, $p=-0.55$, and $q = 35.4$ m, with the upper layer density $\rho \approx 0.4$ g/cc, and $z>0$ for increasing depth.  A different model with a constant firn correction (to sounding propagation times) and no exponential density profile yields shelf depths consistent within errors \citep{ARIANNANIM}.  Equations \ref{nz1} and \ref{nz2} are based on measurements taken at Williams Field near McMurdo station \citep{Barrella,Schytt}.  Given the measured physical delay between pulse and reflection, $\Delta t$, the shelf-depth can be obtained from integrating over the total path length $d$ (Eqs. \ref{d1} and \ref{d2}).  Error propagation yields Eq. \ref{err1}, where $D_f = 67 \pm 10 $ m is the firn depth \citep{DandE2004}.  Figure 2 of the latter reference contains a density profile for the RIS which is consistent with this model.

\begin{align}
\frac{c\Delta t}{2} &= \int_0^{d_{ice}} n(z) dz \label{d1} \\
d_{ice} &= \frac{c\Delta t}{2 n} - \frac{D_f (n_0 - n)}{n} + \frac{q p}{n} (e^{-D_f/q}-1) \label{d2}
\end{align}

\begin{align}
\begin{split}
\sigma_{d,ice} &= \sqrt{\left(\frac{\sigma_t c}{2n} \right)^2+\left(\frac{\sigma_n c\Delta t}{2n^2}\right)^2+k\sigma_{Df}^2} \\
& \approx \frac{c}{2n} \sqrt{\left(\frac{\sigma_n}{n}\right)^2 \Delta t^2 + \sigma_t^2}
\label{err1}
\end{split}
\end{align}

The fractional difference between $n_0$ and $n$ is small, and $\exp(-D_f/q)$ is small, so $k$ turns out to be of order $10^{-2}$.  The term in Eq. \ref{err1} involving $k$ is a factor of 10 below the others so it may be dropped.  For similar reasons, cross-terms involving firn properties and $\sigma_n$ have been neglected.

\section{Experimental Technique}\label{sec:tech}

\begin{figure}
\begin{center}
\includegraphics[trim=2cm 1.0cm 2cm 1.0cm,clip=true,width=80mm]{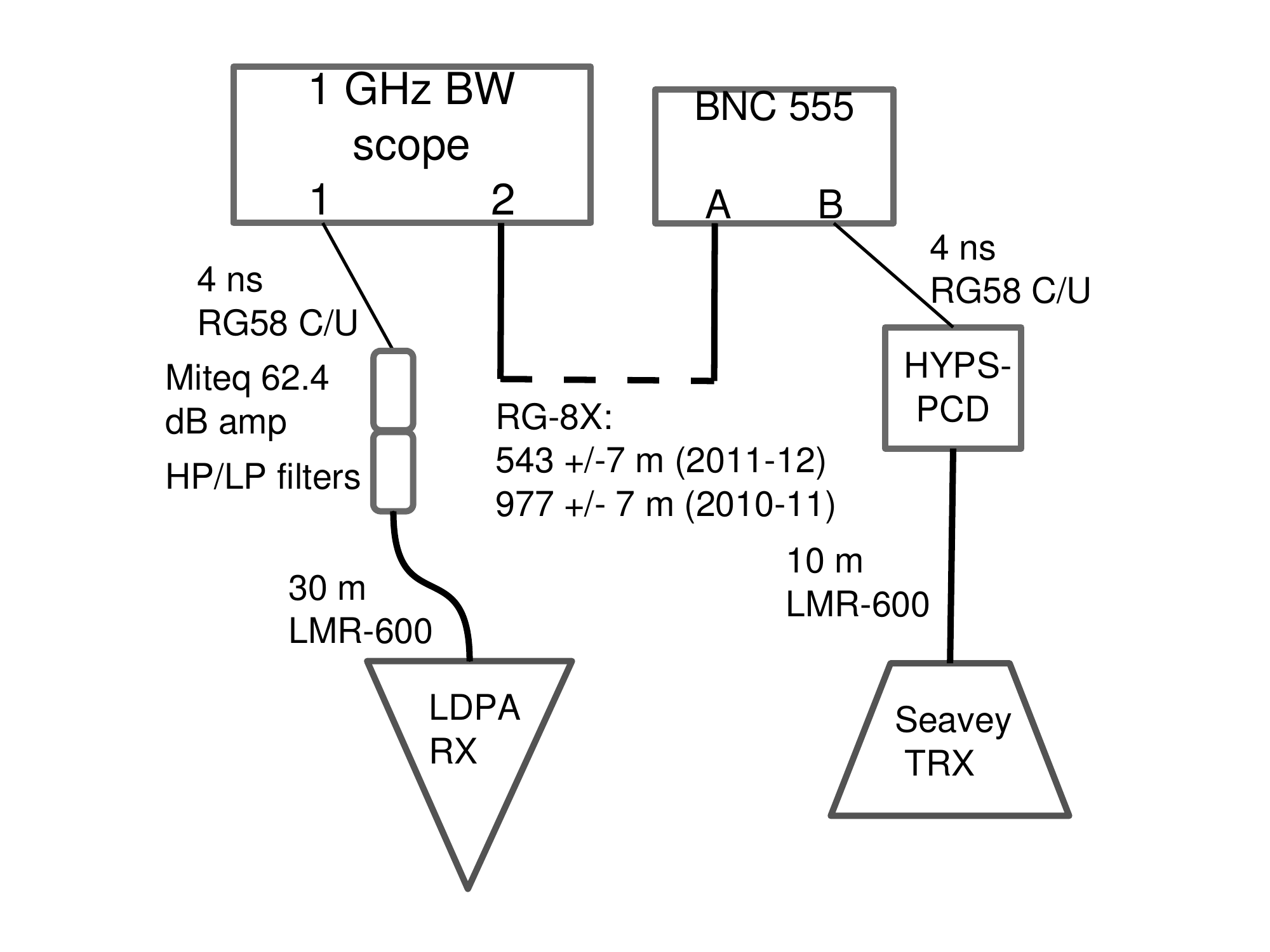}
\end{center}
\caption{\label{setup} The general set-up of the radio-sounding experiments.  Measured and physical time delays are shown in Tab. \ref{tab0}.}
\end{figure}

\begin{figure}
\begin{center}
\includegraphics[trim=0.5cm 0.5cm 0.5cm 0cm,clip=true,width=80mm]{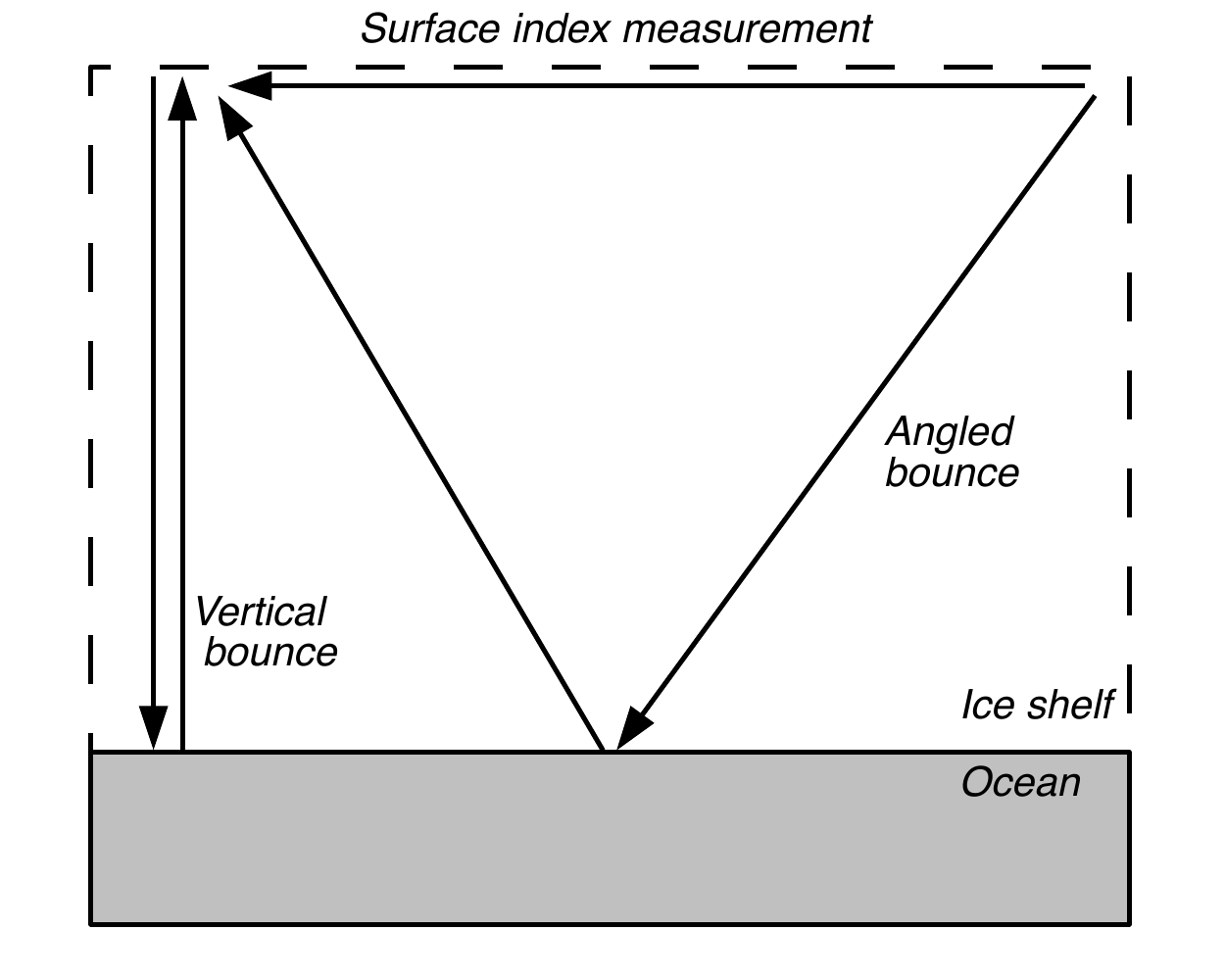}
\end{center}
\caption{\label{bounce}  The vertical and angled bounce tests.  The surface propagation set-up was used to derive the sufrace index of refraction, $n_{surf}$.}
\end{figure}

The experimental set-up is shown in Fig. \ref{setup}.  Figure \ref{bounce} shows the vertical and angled bounces.  To create broad-band RF pulses, a 1 ns wide, 1-2.5 kV pulse was delivered from the HYPS Pockels Cell Driver (PCD) to a transmitting antenna, and the reflection is received by a second antenna.  The PCD and the 1-GHz bandwidth oscilloscope (Tektronix TDS540A in 2010, Agilient HP54832D thereafter) were triggered with a tunable delay generator (Berkeley Nucleonics 555 2-port).  From the programmed delay, reflection time, and relevant cable delays, the shelf-depths were derived for each season.  The RG-8X cable between port A of the BNC 555 and the oscilloscope enabled the introduction of a long baseline in between the antennas.

Voltage standing wave ratio (VSWR) measurements were performed to study antenna transmission in snow.  The VSWR of the transmitting and receiving antennas, in all cases, demonstrate good transmission and reception when buried in the surface snow \citep{Barrella,ARIANNANIM}.  Noise above and below the receiver bandwidth was filtered with MiniCircuits NHP and NLP filters, and amplified by a 62.4 dB Miteq AM-1660 low-noise amplifier (typical noise figure of 1.5 dB).  Signals were attenuated by 3-20 dB where appropriate to remain in the linear regime of the amplifier.

For the 2006 season \citep{Barrella}, the transmitter and receiver were Seavey radio horns used in the ANITA experiment \citep{ANITADesign}, with a bandwidth of [200-1300] MHz.  In the data from the 2010 season, the receiver and transmitter were log-periodic dipole arrays (Create Corp. CLP5130-2N) with a bandwidth of [100-1300] MHz.  The Seavey is a dual polarization quad-ridge horn antenna that has higher gain above 200 MHz than the LPDA.  The LPDA antennas have a wider bandwidth, but stretch the signal in time with respect to the horn \citep{timeconv}.  In the 2011 season, the data was recorded with a Seavey transmitter, and an LPDA receiver.  The 2010 data has been published \citep{Hanson2011,Jordo}.  In this work, the thickness results from 2010 are compared to three new measurements, and a new reflection coefficient and attenuation length analysis are presented.

In the surface test, we measured the pulse propagation time over the $543\pm7$ m baseline, and extracted the surface index of refraction from the speed.  The result was $n_{surf} = 1.29\pm0.02$, and is needed for the boundary conditions in shelf-thickness model.  In the $\emph{vertical}$ $\emph{bounce}$ measurements, where the transmitter and receiver are co-located, the separation in 2006 was typically 9 m.  In 2010 and 2011 the separation was 19 and 23 m, respectively.  This ensures that the receiver is in the far field of the transmitter during calibration.  Comparing vertical bounce soundings to calibration measurements allows derivation of $\langle L \rangle$ assuming a value for $\sqrt{R}$.

The $\emph{angled-bounce}$ measurements are also compared to calibration measurements and vertical bounce cases to measure both $\langle L \rangle$ and $\sqrt{R}$.  Angled signals were captured without having to account for complex ray tracing near the surface.  During angled bounce tests, the transmitter and receiver were angled $45^{\circ}$ downward from horizontal.  For the angled bounce measurements, the 2010 baseline was $977\pm7$ m, and the 2011 baseline was $543\pm7$ m.  The angled bounce measurements in 2010 and 2011 had signal path lengths of $1517\pm8$ m and $1272\pm7$ m, respectively.   The incident angle with respect to normal refracts closer to $30^{\circ}$ when the pulse reaches the ocean, because the firn index $n_{firn} = 1.3$ is smaller than the bulk ice index $n_{ice} = 1.78$

\begin{table}
\setlength\tabcolsep{8pt}
	\begin{center}
		\begin{tabular}{cccc}
			\hline
			\hline
			Year & Vertical/Angled & Ant. (TX/RX) & $G_1G_2$ \\ \hline
			2010 & Vertical & L/L & 1.0 \\ \hline
			2011 & Vertical & S/L & 1.0 \\ \hline
			2011 & Angled & S/L & 0.5 \\ \hline
			2011 & Vertical & L/L & 1.0 \\ \hline
			\hline
		\end{tabular}
	\end{center}
	\caption{\label{tabSetup} The various exerimental configurations used, by year, for the data in this work.  In the third column, S stands for Seavey horn, and L stands for LDPA.}
\end{table}

The Friis equation relates the power received $P_r$ to the transmitted power $P_t$ in a lossless medium at a given wavelength.  For two identical antennas in air, it may be written

\begin{equation}
P_r = \frac{P_t (G_a c)^2}{(4\pi \nu d)^2} = \frac{P_0}{d^2}
\label{Friis}
\end{equation}

In Eq. \ref{Friis}, $G_a$ is the intrinsic gain of the antennas and $\nu$ is the frequency.  $P_r$ and $P_t$ are the received and transmitted power, respectively.  To account for absorption losses and possible losses upon reflection, the Friis equation is modified to

\begin{equation}
P_r = \frac{P_0 R G_1 G_2}{d^2} \exp \left( - \frac{2 d}{\langle L \rangle}\right)
\label{Friis2}
\end{equation}

By convention, the factor of 2 in the exponential means $\langle L \rangle$ refers to electric field, and the reflection coefficient for the power is $R$. The factor $G_1G_2$ accounts for the relative power radiation pattern of the transmitter and receiver (Tab. \ref{setup}).  $G_1$ and $G_2$ are 1 for the vertical bounce measurements, in which the signal is transmitted and received in the forward direction of the antennas.  As the angle at which the signal interacts with the antenna increases, $G_1 G_2$ decreases from 1 according to the antenna radiation patterns.  The radiation patterns have been both simulated and measured \citep{ANITADesign,timeconv}.  Manipulating Eq. \ref{Friis2} gives Eq. \ref{FinalTech1}, the left hand side of which may be plotted vs. path-length $d$ to obtain a line with a slope $-1/\langle L \rangle$, and a constant y-intercept.  The reflection coefficient is treated as a free parameter in the fit.  The error in the left-hand side of Eq. \ref{FinalTech1} is given by Eq. \ref{FinalTech2}.

\begin{align}
f(d) &= \ln \sqrt{\frac{d^2 P_r}{R G_1 G_2}} = -d/\langle L\rangle + const \label{FinalTech1} \\ 
\sigma_f &= \sqrt{\left( \frac{\sigma_d}{d}\right)^2 + \left( \frac{\sigma_{P,r}}{2 P_r}\right )^2} \label{FinalTech2}
\end{align}

\section{Ice Thickness Results}\label{sec:depthresult}

The measured propagation times, both total and corrected for cable delays, are shown in Tab. \ref{tab0}, along with statistical and systematic uncertainties.  Using Eqs. \ref{d1}-\ref{err1}, the times are converted to thickness.  Systematic errors arise from cables, the response time of the pulser, and delay generator precision (see Fig. \ref{setup}).  For the long cables, a conservative $5\%$ error in the propagation time is assumed, because the end-points of the baselines were measured with GPS, with way-point precision of 5 m.  The distance between the 2006 site and the site for the remaining three thickness measurements was approximately 1 km, and the GPS location of the latter site is accurate to within a horizontal uncertainty of $5$ m.

\begin{table}
\setlength\tabcolsep{4pt}
	\begin{center}
		\begin{tabular}{cccccccc}
			\hline
			\hline
			Year & $\Delta t_{meas}$ & $\Delta t_{phys}$ & $\sigma_{stat}$ & $\sigma_{sys}$ & $\sigma_{pulse}$ & $\sigma_{tot}$ & $d_{ice}$ (m) \\ \hline
			2006 & - & 6783 & - & - & - & 10 & $577.5\pm10$ \\ \hline 
			2009 & - & 6745 & - & - & - & 15 & $572\pm6$ \\ \hline
			2010 & 7060 & 6772 & 5.0 & 8.0 & 10 & 14 & $576\pm6$ \\ \hline
			2011 & 6964 & 6816 & 4.0 & 5.0 & 10 & 12 & $580\pm6$ \\ \hline
			\hline
		\end{tabular}
	\end{center}
	\caption{\label{tab0} A summary of total and physical time delays for the various seasons, and calculated shelf thicknesses.  All times are in nanoseconds.  The physical time delay $\Delta t_{phys}$ is the measured delay $\Delta t_{meas}$, with equipment delays subtracted.  The total precision is quoted in the earlier measurements \citep{Barrella,ARIANNANIM}.  The width of the reflected pulse, $\sigma_{pulse}$, is caused by the response of the antennas.}
\end{table}

The reflected pulse from the 2009 set-up was only several mV above noise backgrounds, so the entire pulse width was included as systematic error \citep{ARIANNANIM}.  For the 2011 data, the location and uncertainty in peak voltage oscillations in the reflections were used instead, because the signal was well above backgrounds.  The low-frequency ringing in the reflected data originates from group delay in the LPDA, which is 10 ns at 0.2 GHz (the lowest frequency emitted by the transmitter).  When folded into the timing uncertainties, smaller but comparable errors to 2006 are obtained for the thickness.  Timing uncertainties are lowest in 2006 because both transmitter and receiver were Seavey horns, which have lower group delay than the LPDA.

In general, statistical errors come from Eq. \ref{err1}, with $n=1.78\pm0.02$, and total timing error from Tab. \ref{tab0}.  The magnitude of $\sigma_n$ comes from measurements made at the surface ($n_{surf} = 1.29\pm0.02$) \citep{Jordo}.  Fluctuations in $n$ are largest at the surface, making this a conservative estimate for the bulk ice, and it is similar to previous work \citep{ARIANNANIM,Barrella}.  The total error from 2006 is higher because a larger error on the dielectric constant was used.  The $\emph{mean}$ thickness over all seasons is $d_{ice} = 576\pm8$ m (statistical and systematic added in quadrature).  Errors in Tab. \ref{tab0} other than from the index of refraction are treated as systematic.  A linear fit to the four data points together yields a slope consistent with zero (within errors).  The measurement from 2006 took place 1 km from the location of subsequent seasons, and does not deviate strongly from the mean.

\section{Average Attenuation Length, $\langle L \rangle$, vs. Frequency}\label{sec:lambdaF}

The technique of measuring $\sqrt{R}$ and $\langle L \rangle$ simultaneously is more challenging than assuming a constant $\sqrt{R}$ and comparing the raw power spectra of only vertical bounce data and calibration data.  Assuming a uniform reflection coefficient, with respect to frequency, assumes that the reflecting surface is dominated by specular reflection, rather than diffuse reflection.  As long as the first few Fresnel zones $D_m$ of the transmitted pulse are not significantly larger than the horizontal correlation length $L_C$ of roughness features along the shelf base, then the effect of the vertical roughness scales on the reflection coefficient is avoided \citep{Peters}.  Prior data collected at two locations on the Ross Ice Shelf, near Moore's Bay, indicate horizontal correlation lengths $L_C=12.5$ m and $L_C=27.5$ m at the two sites \citep{Neal82}.  The glaciological Fresnel zone equation, for an observation point a distance $h$ above the snow surface, with a shelf thickness of $z$, shelf index of refraction $n$, Fresnel zone number $m$, and an in-air wavelength $\lambda$ is

\begin{equation}
D_m \approx \sqrt{2 m \lambda \left(h+\frac{z}{n}\right)} \label{Fresnel}
\end{equation}

The approximation arises from the small angle approximation, and is sound because the Fresnel zones are small compared to $z$.  The measurements take place at the surface, so $h=0$.  Using $n=1.78$, $\lambda = 3$m, and the measured shelf thickness, Eq. \ref{Fresnel} gives $D_1 = 10-40$ m, for the bandwidth.  Vertical rms fluctuations at the ocean/ice surface were reported to be be 3 cm and 10 cm for two sites, spread out over a typical length scale of $L_C$.  Vertical height fluctuations of 10 cm and 3 cm spread out over 12.5 m and 27.5 m, respectively, means that specular reflection is a good approximation for this bandwidth \citep{Neal82}.  The attenuation lengths derived assuming constant $\sqrt{R}$ are revised in the next section, to account for reflection loss ($\sqrt{R}<1.0$).

Consider the calibration pulse, $V_C$, the vertical bounce pulse, $V_{ice}$, and the depth-averaged attenuation length vs. frequency, $\langle L(\nu) \rangle$, all at a frequency $\nu$:

\begin{align}
V_{C}(\nu) &= V_{0}/d_{C} \label{V1} \\
V_{ice}(\nu) &= \frac{V_{0}}{d_{ice}} \exp \left( -\frac{d_{ice}}{\langle L(\nu) \rangle}\right) \label{V2} \\
\langle L(\nu) \rangle &= \frac{d_{ice}}{\ln ( (V_{C}(\nu)d_{C}) / (V_{ice}(\nu)d_{ice}) )}
\label{V3}
\end{align}

Because the surface of the firn is snow, with a density of 0.4 g/cc, and an index of refraction $n=1.3$, the reflection coefficient (for power) between air and snow is $\approx 0.02$, so potential interference from surface reflections are not expected to modify Eq. \ref{V1}.  The antennas were placed at the maximum height allowed by the cables and other equipment (1.5 m), and this calibration was compared to the case with the antennas buried in snow slots.  Because of the low snow density, dielectric absorption is negligible over the calibration distances (23 m).  The antenna calibrations produced similar waveforms with the antennas lowered in snow.  The waveform amplitude increases when LPDAs are in the snow, due to the shift in the lower cutoff frequency by the index of refraction.  This effect is confirmed in NEC antenna simulations, and VSWR data \citep{timeconv}.

\begin{figure}
\begin{subfigure}[b]{0.3\textwidth}
\includegraphics[width=80mm]{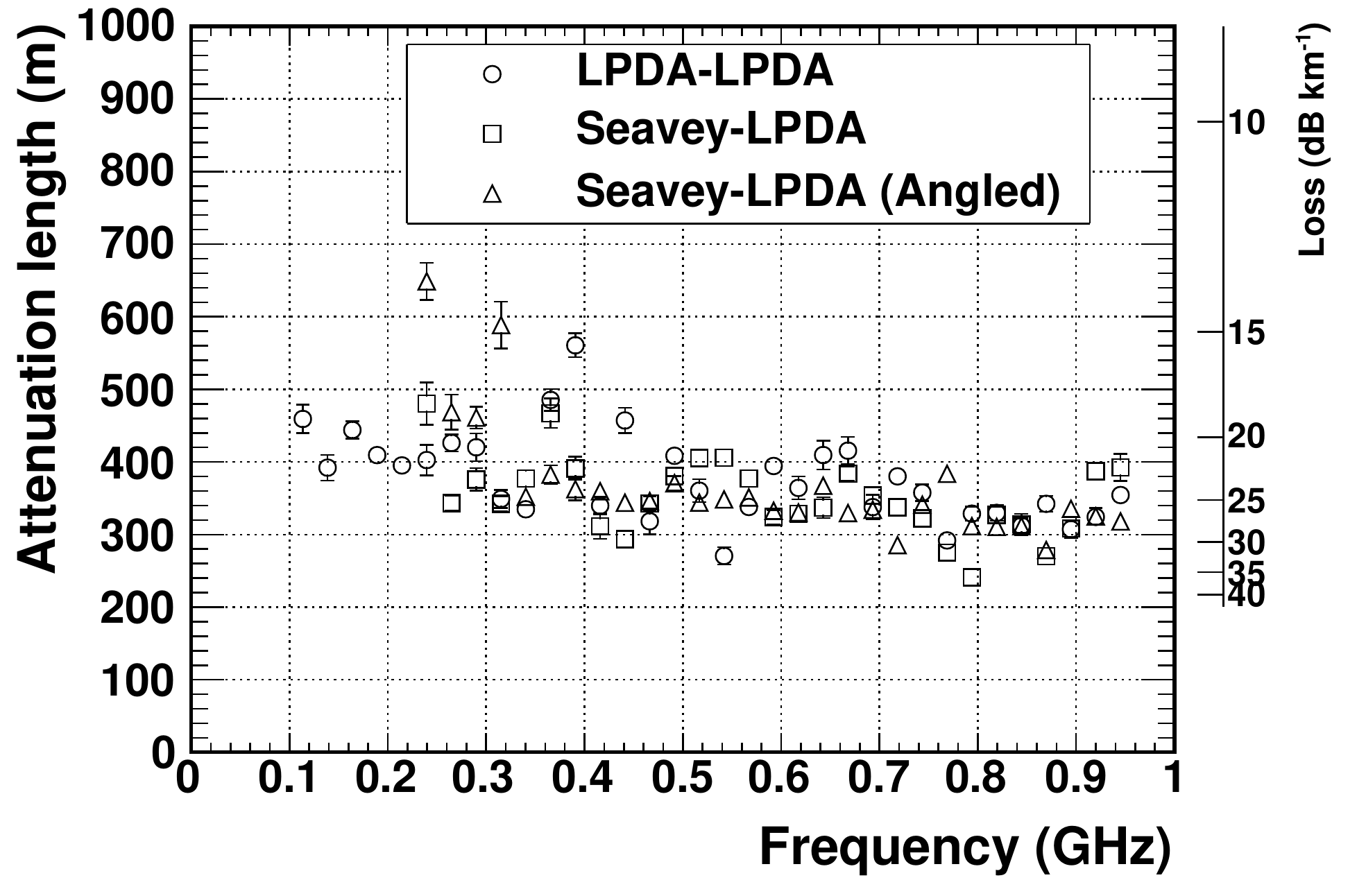}
\end{subfigure}

\begin{subfigure}[b]{0.3\textwidth}
\includegraphics[width=80mm]{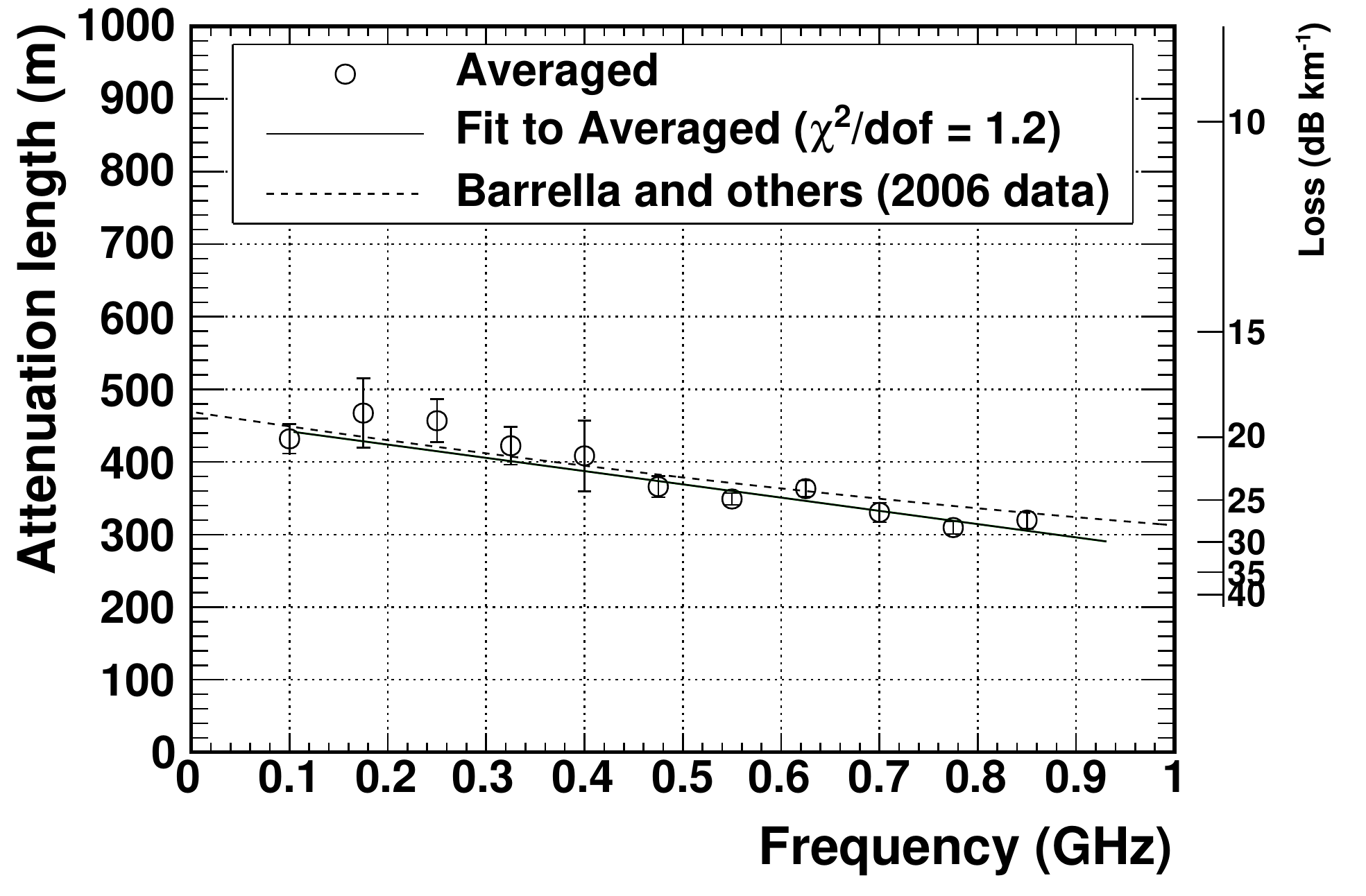}
\end{subfigure}
\caption{\label{lambdaF1} Top: The depth-averaged attenuation length vs. frequency, with standard deviations from error propagation in Eq. \ref{V3}.  The attenuation length is converted to englacial loss (dB/km), at right.  Bottom: The data from Fig. \ref{lambdaF1}, averaged into 75 MHz bins, with standard deviations.  The  linear fit has a $\chi^2/dof=1.2$, a slope of $-180\pm40$ m/GHz , and an offset of $460\pm20$ m.  The dashed line is a fit to prior data taken 1 km from our site \citep{Barrella}.}
\end{figure}

The 2011 data, is shown in Fig. \ref{lambdaF1} (top).  In Eqs. \ref{V1}-\ref{V3}, the voltages are defined like $V \propto \sqrt{P(\nu)}$, where $P$ is the measured power at the frequency $\nu$.  The antenna impedance is the same for the calibration and bounce studies, making it irrelevant in the ratio in Eq. \ref{V3} \citep{timeconv}.  The 2011 power spectra begin at the low-frequency cutoffs of the transmitter type (200 MHz for the Seavey, and 100 MHz for the LPDA).  The englacial loss in dB/km is also shown (Eq. \ref{stupidAssReviewWhoDoesntHaveACalculator}).

The 2011 data extend to 0.850 GHz, where the signal to noise ratio is close to 1.0, and the error-bars are the standard deviation from error propagation in Eq. \ref{V3}.  About 10 m of the error is due to uncertainty in the shelf thickness, and 10 m is due to uncertainty in the power spectrum.  Data above 0.850 GHz appear to be rising due to noise floor contributions.  Also, systematic fluctuations in the vertical bounce power spectra lead to systematic fluctuations in $\langle L(\nu) \rangle$.  Systematic errors arise from differences in system frequency response after changing the transmitter location and type, and reflections within the system.  The angled bounce data at 0.240 and 0.315 GHz in particular are systematically high.  The Seavey transmitter was placed in a snow cavity rather than fully buried for the angled test, which can lead to cavity resonance effects.

The frequency resolution has been maximized in Fig. \ref{lambdaF1} (top), with no window function.  A higher resolution extends the upper frequency limit by avoiding folding noise into the highest frequency bins.  The correction for potential angular dependence of the reflection coefficient only applies to the angled bounce data ($\sim 4$ m).  In Fig. \ref{lambdaF1} (bottom), the data are averaged into 0.075 GHz bins, with a linear fit.  The best-fit slope is $-180\pm40$ m/GHz, the best-fit offset is $460\pm20$ m (95\% confidence level, $\chi^2/dof=1.2$).  Data above 0.850 GHz has been neglected in the average and fit shown in Fig. \ref{lambdaF1}, however, the $\chi^2/dof$ only increases to 1.8 if it is included.  As in Fig. \ref{lambdaF1} (top), the averaged attenuation length is converted to dB/km on the right-hand y-axis using Eq. \ref{stupidAssReviewWhoDoesntHaveACalculator}.

Despite the systematic fluctuations, the fit to the data in Fig. \ref{lambdaF1} is in close agreement with the quadratic fit to the data from 2006 \citep{Barrella}.  In the publication of the 2006 data, the reflection loss was assumed to be 0 dB.  If a lower value is assumed (see below), the attenuation length $\emph{increases}$, because the returned voltage per unit frequency in Eq. \ref{V3} must remain constant.  The level of systematic variation in $\sqrt{R}$ shown below would also generate $\approx5$\% systematic uncertainty in $\langle L \rangle$, but only to increase it.  The 2006 and 2011 agree, even though the measurements were made 1 km apart.  The area of Moore's Bay near Minna Bluff is far from any zones of high glacial velocity that could cause depth or basal reflection variations, and crevasses have not been observed in the area, so the ice is expected to be relatively uniform.

\section{The Basal Reflection Coefficient}\label{sec:lambdaR}

The 2006 season $\langle L \rangle$ results were derived from vertical bounce measurements assuming $\sqrt{R} = 1.0$.  Using the path lengths derived from shelf thickness, and the measured power spectra of the calibration, vertical bounce, and angled bounce reflections, $\sqrt{R}$ can be derived separately from the attenuation length.  The errors in $\sqrt{R}$ arise from propagating errors in path length (from thickness, and geometry) and returned power through Eq. \ref{FinalTech1}.

The three tests (calibration, vertical, and angled bounce) serve as three measurements of $f(d)$ for different values of the path length $d$, given the free parameter $\sqrt{R}$.  The measurements are compared to the linear model $f_{model} = -d/\langle L \rangle +f_0$, which is scanned through $(\sqrt{R},\langle L \rangle)$ parameter space.  The y-intercept is irrelevant to the physics, coming from the linear fit upon each iteration.  (The overall power at a given frequency is relative to the calibration pulse power).  Each iteration produces a $\chi^2$ value, and $(\sqrt{R},\langle L \rangle)$ were scanned until a global minimum was reached at each frequency.

\begin{figure}
\begin{center}
\includegraphics[trim=0.5cm 5cm 1cm 5cm,clip=true,width=85mm]{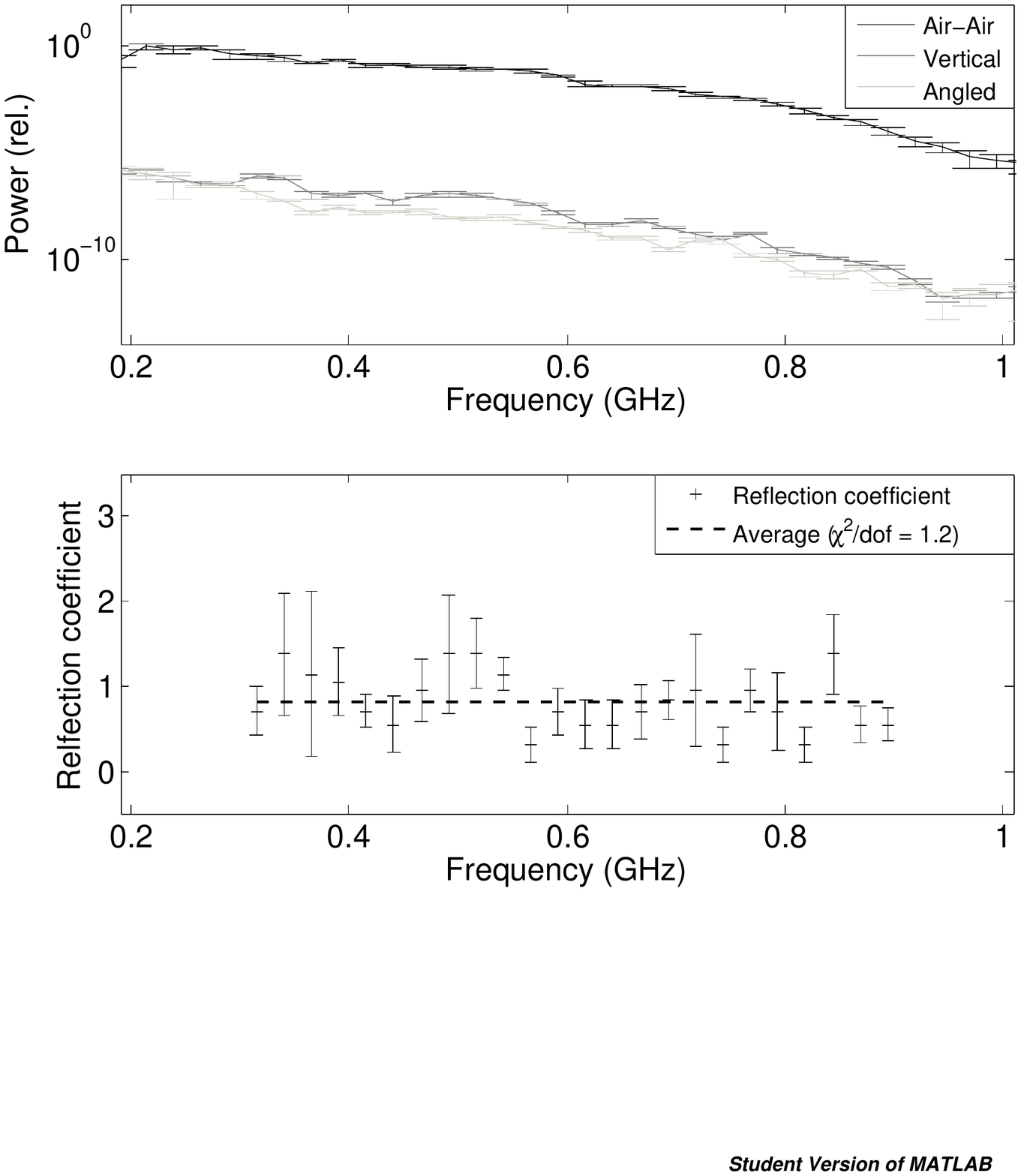}
\end{center}
\caption{\label{lamdaRFig2} The electric-field reflection coefficient, $\sqrt{R}$, versus frequency.  The three power spectra correspond to three measurements: a surface power calibration (black), vertical bounce (dark grey) and angled bounce (light grey) cases.  The three measurements at each frequency determine a reflection coefficient through a linear fit to Eq. \ref{FinalTech1}, with errors from Eq. \ref{FinalTech2} attributed to $\sqrt{R}$.}
\end{figure}

\begin{figure}
\begin{center}
\includegraphics[width=80mm]{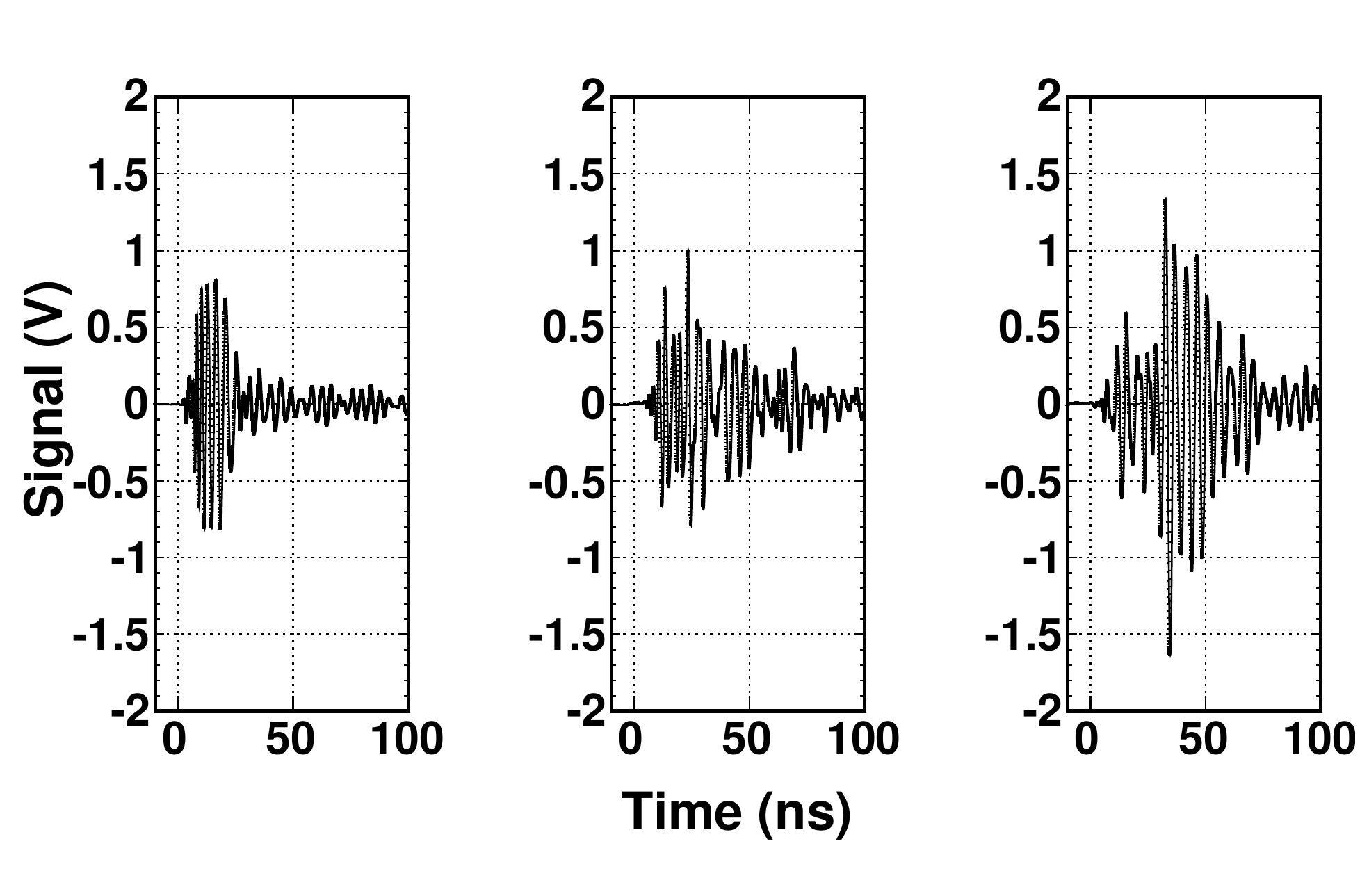}
\end{center}
\caption{\label{time2} Typical waveforms from 2011, averaged over 100 triggers. (Left): the calibration pulse.  (Middle): the vertical bounce reflection.  (Right): the angled bounce reflection.  The vertical and angled bounce data have been amplified by the 60 dB Miteq amplifier.  The data has been scaled to account for in-line attenuators (the data was kept within the amplifier linear range).}
\end{figure}

The averaged power spectra of the time-dependent waveforms are shown in Fig. \ref{lamdaRFig2} (top).  The spectra are constructed from averaging the modulus-squared of the FFT of the time-dependent signals, and plotted relative to the maximum calibration power.  The error-bars are the error in the mean for each bin.  Examples of waveforms from which these power spectra are derived are shown separately in Fig. \ref{time2}.  For all recorded waveforms, a sampling rate of 5 GHz was used on the 1-GHz bandwidth oscilloscope.  The spectra in Fig. \ref{lamdaRFig2} have a frequency resolution of 0.025 GHz.

An analysis of the 2010 data for $\sqrt{R}$ vs. $\langle L \rangle$ has been shown in \citep{Hanson2011,Jordo}. The basic results were 480 m $\leq \langle L \rangle \leq 510$ m (17-18 dB/km), and $0.72 \leq \sqrt{R} \leq 0.88$ (1.1-2.8 dB loss), for the average attenuation length and reflection coefficient (68\% confidence level).  The set-up (Fig. \ref{setup}) demonstrated good transmission through surface snow for frequencies below 0.180 GHz that season, and the LDPA lower limit in the snow is 0.080 MHz.  The index of refraction of snow extends the LPDA response to 0.080 GHz from a lower limit of 0.105 GHz \citep{timeconv}.  A shorter angled bounce baseline (543$\pm$7 m) was chosen for the 2011 season, relative to the prior year, to boost signal at higher frequencies, however the snow absorption effect was not observed in 2011.

Figure \ref{lamdaRFig2} (bottom) shows the $\sqrt{R}$ results from the 2011 season.  The baseline sets the path length difference between the angled and vertical cases, introducing a trade-off.  A shorter baseline causes the attenuation length to become large compared to the difference in path-length between the angled and vertical bounce tests ($\approx130$ m in 2011).  At low frequencies, the difference in power loss between vertical and angled cases becomes smaller than the errors in the power spectra (about 3 dB at 0.300 GHz).  Alternatively, the baseline for the angled bounce can be increased, which increases the path length difference between angled and vertical bounces.  While this increases the low-frequency precision, the high-frequency precision suffers due to increased absoprtion in the angled bounce data.  The vertical and angled signal power are equal within statistical errors up to 0.300 GHz in Fig. \ref{lamdaRFig2} (top), but differences in the vertical and angled power are measurable up to 0.850 GHz.

Figure \ref{lamdaRFig2} (bottom) shows $\sqrt{R}$, with statistical errors from the fit, at each frequency.  The errors are conservative, in that all the deviation from a perfect linear model (Eq. \ref{FinalTech2}) is attributed to error in $\sqrt{R}$.  The mean is $\sqrt{R}=0.82\pm0.07$ (-1.7 dB), and a flat model at this mean has a chi-squared result of $\chi^2/dof=1.2$.  Despite fluctuations in the data, no data point deviates above the physical region by much more than one standard deviation.  At each point in the bandwidth where the gap between the reflected and calibration spectra randomly decreases, the parameter $\sqrt{R}$ must flucuate upwards to produce a linear fit to $f(d)$.  These data have larger error bars, because the deviation in $f(d)$ from a linear model is larger for those bins.  An unweighted flat line fit to the data is shown; this fit produces the same results, within statistical uncertainties, as a weighted fit (that de-emphasizes the points with large errors).  If the bins with $\sqrt{R}>1$ are ignored in the fit, then the mean decreases by 20\% ($\sqrt{R}=0.6$).

Ignoring the data above 1.0, however, would raise the attenuation length results, because the total power loss must be conserved (Tab. \ref{RL}).  As the attenuation length has the stronger impact on the design of the future ARIANNA detector, relative to the reflection coefficient, it is important to be conservative with regard to the attenuation length in Tab. \ref{RL} \citep{Kam}.  The origin of the roughness in the reflected spectra is likely noise interference, since the signal to noise ratio is lower than the calibration study.

Finally, knowledge of the basal reflection coefficient allows the correction of the attenuation length numbers in Fig. \ref{lambdaF1} to more realistic values.  If $\langle L_0 \rangle$ is the measured attenuation length, assuming $\sqrt{R}=1.0$, then the actual attenuation length $\langle L \rangle$ can be expressed as

\begin{equation}
\frac{\langle L \rangle}{\langle L_0 \rangle} = \left(1+\frac{\langle L_0 \rangle}{2 d_{ice}} \ln R \right)^{-1}
\end{equation}

Using the $\langle L_0 \rangle$ values from Fig. \ref{lambdaF1}, Tab. \ref{RL} shows the $\langle L \rangle$ results for the mean value of $\sqrt{R}=0.82\pm0.07$, versus frequency.  Table \ref{RL} also shows the imaginary part of dielectric constant, derived from $n''$, via $\langle L \rangle^{-1} = n''k$, where $k$ is the free-space wavenumber.  Assuming that $\tan\delta\ll1$, the expression $\epsilon'' = 2n''\sqrt{\epsilon'}$, with $\sqrt{\epsilon'}=1.78$, relates the two quantities.

The $\epsilon''$ results are in agreement with an earlier low-­frequency projection for typical ice ­shelf temperatures \citep{Mat}.  The Debye model predicts $\epsilon'' \propto \nu^{-1}$ for frequencies below 2 GHz, and the $\epsilon''$ data follow this trend.  The quantity $\nu\tan\delta$ is expected to be small and constant for a simple dielectric material, and Tab. \ref{RL} also displays this quantity in the final column, which agrees with an estimate from analysis of the 2006 data \citep{Barrella}.  Although $\nu\tan\delta$ varies with frequency, this variation is such that no measurement is more than one standard deviation ($0.2\times 10^{-4}$) from the mean ($1.37\pm0.06$).

\begin{table}
\setlength\tabcolsep{3pt}
	\begin{center}
		\begin{tabular}{cccccc}
			\hline \hline
			$\nu$ (GHz) & $\langle L_0 \rangle$ (m) & $\langle L \rangle$ (m) & (dB/km) & $\epsilon'' \times 10^3$ & $\nu\tan\delta \times 10^4$ \\ \hline
			0.100 & 432 & 449 & 19.3 & $3.8$ & $1.2$ \\ \hline
			0.175 & 467 & 487 & 17.8 & $2.0$ & $1.1$ \\ \hline
			0.250 & 457 & 476 & 18.2 & $1.4$ & $1.1$ \\ \hline
			0.325 & 422 & 438 & 19.8 & $1.2$ & $1.2$ \\ \hline
			0.400 & 408 & 423 & 20.5 & $1.0$ & $1.3$ \\ \hline
			0.475 & 366 & 378 & 23.0 & $0.95$ & $1.4$ \\ \hline
			0.550 & 349 & 360 & 24.1 & $0.86$ & $1.5$ \\ \hline
			0.625 & 363 & 375 & 23.2 & $0.72$ & $1.4$ \\ \hline
			0.700 & 331 & 341 & 25.5 & $0.71$ & $1.6$ \\ \hline
			0.775 & 310 & 319 & 27.2 & $0.69$ & $1.7$ \\ \hline
			0.850 & 320 & 329 & 26.4 & $0.61$ & $1.6$ \\ \hline \hline
			Ave. & $380\pm16$ & $400\pm18$ & $22\pm1$ & $1.3\pm0.3$ & $1.37\pm0.06$ \\ \hline
			\hline
		\end{tabular}
	\end{center}
	\caption{\label{RL} Summary of dielectric parameters.  The first column is the frequency, $\nu$ (GHz), followed by the attenuation lengths, which are un-corrected ($\langle L_0 \rangle$) and corrected ($\langle L \rangle$) for $\sqrt{R}=0.82\pm0.07$.  The fourth column is $\langle L \rangle$, expressed in (dB/km).  The imaginary part of the dielectric constant, $\epsilon''$, is shown in the fifth columns.  The final column shows $\nu\tan\delta$ (in units of GHz).  The typical error on the quantity $\nu\tan\delta$ is $0.2\times 10^{-4}$.}
\end{table}

\section{Polarization Measurements}\label{sec:pol}

The $\sqrt{R}$ result shows that little power is lost from the basal reflection.  In this section, we assess potential losses by scattering or rotation of the linearly polarized signal.  For any non-ideal linearly polarized antenna system, a small amount of power can leak into the cross-polarized channel.  Significant transfer of power into the cross-polarized direction would indicate polarization rotation in the ice, and bias the attenuation length results.  Birefringence and surface roughness effects at the water-ice interface at the bottom of the ice shelf are expected to generate power in the cross-polarized direction.

To quantify the polarization rotation, the cross-polarization fraction, $F_{ice}$, was measured in the vertical bounce configuration, and compared to $F_{air}$.  $F$ is defined in Eq. \ref{F}, where $P_{\perp}$ and $P_{||}$ refer to the measured power in the cross-polarized and co-polarized direction with respect to the linear polarization of the transmitter, at a given frequency.

\begin{equation}
F = \frac{P_{\perp}}{P_{\perp}+P_{||}}
\label{F}
\end{equation}

The leakage between co-polarized and cross-polarized channels is expected to be low across the bandwidth, but difficult to observe at high frequencies.  Cross-polarized signals are weaker than co-polarized, and the vertical bounce data in the cross-polarized state is subject to noise interference above 0.4 GHz.  The intrinsic transfer into the cross-polarized direction of a specified antenna pair was estimated by facing the transmitter toward the receiver in air.  $F_{air}$ is computed from the power observed between co-polarized and cross-polarized orientation of the receiver.  The results of this study are shown in the third column of Tab. \ref{crossP}.  It was verified that the snow surface 1.5 m below the antennas scatters back a negligible amount of power.

\begin{table}
\setlength\tabcolsep{15pt}
\begin{center}
\begin{tabular}{ccc}
\hline \hline
Frequency (GHz) & $F_{air}$ & $F_{ice}$ \\ \hline
0.175 & $0.06\pm0.02$ &  $0.08\pm0.05$ \\ \hline
0.200 & $0.04\pm0.01$ & $0.01\pm0.01$ \\ \hline
0.225 & $0.04\pm0.02$ & $0.02\pm0.01$ \\ \hline 
0.250 & $0.02\pm0.01$ & $0.01\pm0.01$ \\ \hline
0.275 & $0.02\pm0.01$ & $0.02\pm0.01$ \\ \hline
0.300 & $0.02\pm0.01$ & $0.07\pm0.04$ \\ \hline
0.325 & $0.01\pm0.005$ & $0.03\pm0.01$ \\ \hline
0.350 & $0.04\pm0.01$ & $0.08\pm0.03$ \\ \hline
0.375 & $0.02\pm0.01$ & $0.11\pm0.07$ \\ \hline
0.400 & $0.03\pm0.01$ & $0.22\pm0.09$ \\ \hline
\hline
\end{tabular}
\end{center}
\caption{\label{crossP}  A comparison of cross-polarization fraction measurements versus frequency.}
\end{table}

$F_{ice}$ was obtained from the vertical bounce set-up, with a Seavey transmitter and LPDA receiver.  The Seavey antenna transmits very little power below 0.175 GHz, and the cross-polarized signal is weaker than the co-polarized signal, limiting $F_{ice}$ results to frequencies below 0.4GHz.  These measurements are shown in column 4 of Tab. \ref{crossP}.  This data can be compared to measurements taken in 2010, in which $F_{ice}$ and $F_{air}$ were shown to agree at 0.1 GHz with a LPDA transmitter and LPDA receiver at the same location as the 2011 measurements \citep{Hanson2011}.  A comparison of $F_{air}$ and $F_{ice}$ reveals no excess power in the cross-polarization direction, with the possible exception of data at 0.400 GHz, which shows a 2$\sigma$ deviation from intrinsic antenna effects.  This data does not confirm the $F_{ice}$ analysis of the 2006 data, which showed $F_{ice} = 0.7$ at 0.4 GHz.

\section{Discussion}

The data is in agreement with independent analyses and models.  A study from Greenland found the total transfer function of the Greenland ice sheet, and models the different contributions from basal reflection and attenuation \citep{Paden}.  A reflection coefficient (for power) of -37 dB is reported for the NGRIP2 location, and ice absorption of $\approx$ 56 dB.  Given the depth of 3.1 km, a loss rate of $\approx$ 9.0 dB/km is obtained.  (The Greenland study was limited to 0.11-0.5 GHz).  The upper half of the Greenland ice sheet is colder than Moore's Bay, lowering the attenuation rate through temperature dependence of $\epsilon''$.  The reflection coefficient from that study (-37 dB) is much smaller than that of Moore's Bay.  However, other authors have estimated it to be higher \citep{Avva,Bamber}, with an absorption rate of 9.2 dB/km, conservatively assuming no reflection loss (attributing all loss to absorption).  The Greenland site also exhibits a frequency dependence that produces a change of 8.5 dB/km over the bandwidth (0.11-0.5 GHz).  The slope of the loss rate versus frequency is therefore $8.5/(0.55-0.11) \approx 22$ dB/km/GHz.  The corresponding number for the ARIANNA site is 9.3 dB/km/GHz, from Tab. \ref{RL}.

Another study presents models for ice absorption across the entire Antarctic continent, given an array of inputs, such as temperature and chemistry data \citep{Mat2}.  This expansive study presents results for shelf and sheet depth across the continent, and the portion depicting the Ross Ice Shelf, near the ARIANNA site, is in agreement with our thickness measurements.  The RIS depth is peaked at 500 m in this references' model, and we find $576\pm8$ m.  The inputs to this model indicate that the Ronne Ice Shelf has smaller absorption rates (in dB/km) than the Ross Ice Shelf, which leads to a double-peaked distribution of loss rates, with one peak near 12.5 dB/km, and the other near 22.5 dB/km.  The ARIANNA site average absorption rate is within one standard deviation of the mean for the entire distribution (15.1$\pm$6.2 dB/km), and is in agreement with the second peak in the distribution of loss rates, corresponding to the Ross Ice Shelf.

Finally, a study of the Ross Ice Shelf at 2 MHz reveals large-scale thickness uniformity in the shelf, up to 40 km from the grounding line of the glaciers flowing into the shelf \citep{MacGregor}.  The measurements are obtained from basal echoes with travelling transmitters and receivers at the surface.  In some cases, multiple echoes are observed, corresponding to multiple round trips made by the signal, from surface to base.  This technique provides excellent constraints on the thickness and absorption rate.  Specifically, this study shows that our depth measurement is typical for large expanses of ice, a key requirement for large-scale ground-arrays in neutrino detectors.

\section{Conclusion}

During the 2011-12 Antarctic season, radio echo-sounding measurements were performed in Moore's Bay with high-voltage broad-band RF pulses in the 0.1-0.850 GHz bandwidth, to understand the dielectric properties of the ice shelf.  The shelf thickness was determined from the total propagation time to be 576$\pm$8 m.  The echo-soundings revealed depth-averaged attenuation lengths well-fit by the linear function  $\langle L(\nu) \rangle = (460\pm20)-(180\pm40)\times\nu$ m (19.3-26.4 dB/km), where $\nu$ is the frequency (GHz).  The $\chi^2/dof$ of this linear fit to the combination of multiple data sets was 1.2, with 9 degrees of freedom.  The fit is consistent with prior measurements \citep{Barrella}, and the functional dependence is compatible with theoretical expectations \citep{Somaraju,Mat}.

Vertical echo-soundings were compared to echo-soundings with a $543\pm7$ m baseline between transmitter and receiver, which allowed independent measurement of the basal reflection coefficient, found to be $\sqrt{R}=0.82\pm0.07$ (-1.7 dB).  The slope of $\sqrt{R}$ versus frequency is consistent with a flat-mirror approximation.  The average value of $\sqrt{R}$ is consistent with earlier studies performed at lower frequencies \citep{Neal79}.  The short duration of the observed pulses (90\% of the power contained within 100ns) preclude significant multi-path effects.  The Fresnel zones of the pulses at the shelf base are not significantly larger than measured horizontal correlation lengths.  After correcting attenuation lengths for the effect of $\sqrt{R}$ on returned power, dielectric quantities like $\epsilon''$ and $\nu\tan\delta$ were derived.  The results for $\epsilon''$ and $\nu\tan{\delta}$ agree with theoretical expectations \citep{Mat}.
  
Finally, the fraction of scattered power by the ice into the cross-polarized direction, $F_{ice}$, for is less than 10\% (0.100-0.400 GHz), compatible with the fraction of power due to intrinsic limitations of the transmitting and receiving antennas.  Both the large value of $\sqrt{R}$ and the small value of $F_{ice}$ suggest that the bottom surface of the Ross Ice Shelf at Moore's Bay is smooth.  The measurements of $F_{ice}$ do not demonstrate any significant features below 0.400 GHz, where cross-polarized power is noise-limited.  This result, combined with the measured field attenuation length at frequencies between 0.100-0.850 GHz, suggest that the Moore’s Bay region of the Ross Ice Shelf will be an excellent medium for the ARIANNA high energy neutrino project. 

\section{Acknowledgements}

We wish to thank the staff of Antarctic Support Contractors, Lockheed, Raytheon Polar Services, and the entire crew at McMurdo Station for excellent logistical support.  This work was supported by generous funding from the Office of Polar Programs and Physics Division of the US National Science Foundation, grant awards ANT-08339133, NSF-0970175, and NSF-1126672.  In 2010, additional funding was provided through the Department of Energy under contract DE-AC-76SF-00098.  Finally, we thank Professor David Saltzberg for comments and suggestions throughout the expeditions and analysis.

\bibliography{igs}
\bibliographystyle{igs}

\end{document}